\documentstyle[twoside,fleqn,mssl_workshop,psfig]{article}

\newcommand{\etal}{{\it et al.} }        
\newcommand{\xmm}{{\it XMM-Newton} }     
\newcommand{\chandra}{{\it Chandra} }    

\begin{document}

\title{New insights on chemical evolution of galaxies: \xmm\
  observations of M82}

\author{P. Ranalli\thanks{JSPS fellow.}
\address{Cosmic ray laboratory, RIKEN, 2-1 Hirosawa, Wakoshi, Saitama,
351-0198 Japan},
L. Origlia$^{\rm b}$, A. Comastri\address{INAF--O.A.Bologna, via C. Ranzani 1, 40127 Bologna, Italy}
R. Maiolino \address{INAF--O.A.Arcetri, Largo E. Fermi 5, 50125 Firenze, Italy}
}
\begin{abstract}
  We report on the ongoing data analysis of a very deep ($\sim 100$
  ks) \xmm observation of the starburst galaxy M82.  We show some
  details of data analysis and a few results from spatially-resolved
  spectroscopy with the EPIC cameras.  Since M82 is a bright object
  with a complex spectrum, the data reduction of such a deep observation
  has posed many challenges both about the involved astrophysical
  processes and the available data analysis techniques.
  Vertical (with respect to the galaxy plane) abundance gradients are
  discovered. The hints for an under-abundance of Oxygen stemming from
  our previous study are confirmed. The hot X-ray emitting gas is
  shown to have a multi-temperature distribution.
\end{abstract}

\maketitle

\section{Introduction}
The signature of the star formation (SF) history of a galaxy is
imprinted in the abundance patterns of its stars and gas.  Determining
the abundance of key elements released in the interstellar medium
(ISM) by stars with different mass progenitors and hence on different
time scales, will thus have a strong astrophysical impact in drawing
the global picture of galaxy formation and evolution \cite{mwi97}.
It also offers the unique chance of directly witnessing the enrichment
of the ISM \cite{mc94}.  Metals locked into stars give a picture of
the enrichment just prior to the last burst of SF, while the hot gas
heated by SNe~II explosions and emitting in the X-rays should trace
the enrichment by the new generation of stars.

We have started a project to measure the detailed chemical abundances in a
sample of starburst galaxies, for which we obtained high resolution
infrared (J and H band) spectra with the 3.6 m Italian Telescopio
Nazionale Galileo (TNG) and with the ESO VLT, and both proprietary and
archival data from the \xmm and \chandra missions.  Our sample
comprises M82, NGC253, NGC4449, NGC3256 and the {\em Antennae},
sampling two orders of magnitude in star formation, as it ranges from
the 0.3 $M_\odot$/yr of NGC4449 to the $\sim 30 M_\odot$/yr of the
Antennae and NGC3256.

Preliminary results for M82 were achieved with the available \xmm
archival data, and hinted for a confirmation of the expected scenario
in which the gaseous component has a higher content of
$\alpha$-elements than the stellar one, and a similar content of Fe
\cite{or04}.  However, some new issues were posed, since we found a
very low abundance of O and Ne with respect to other $\alpha$-elements
(e.g., O/Mg $\sim 0.2$, Ne/Mg $\sim 0.3$) in the hot gas present in
the central ($<\sim 1$ kpc) regions of M82, which could not be
satisfactorily explained. For this reason, we were granted a deeper
observation of M82. A report on our preliminary analysis appeared in 
\cite{m82escorial06}. Some issues concerning the data analysis
of complex and well-sampled spectra by standard ``black box'' tools,
such as XSPEC, the Mekal and APEC models, etc.\ are presented in this 
contribution. The main paper is currently in preparation.

\section{EPIC data reduction at the limits of standard tools}

M82 is a bright galaxy with an ongoing starburst and a complex
spectrum. Thus, the data reduction of such a deep observation (100 ks)
has proved difficult and time-consuming, posing many challenges about
the involved astrophysical processes and the available software
apparatus.

The EPIC data were analysed with the SAS software, version 6.5.0.
After screening for background flares, about 73 ks of data were
accepted for analysis.  We began our analysis from the central region
of M82, by extracting spectra from a circle with diameter $1^{\prime}$
and centred on the coordinates 09:55:51 and $+$69:40:39.  Background
spectra were extracted from the blank-sky data files
\cite{birmblanksky} after normalisation to the background levels
observed in the M82 data.

The largest emission from hot plasma is found in the centre of M82.
However, many point sources are also present in the same region.  To
choose a model for the point sources, we analysed the spectrum in the
3-8 keV band, and we found that it can be well described by a power
law with photon index $\Gamma=1.60_{-0.03}^{+0.04}$. This power law
has been always included in the models discussed below regarding the
centre of M82.  A Fe thermal line is also detected at $6.66\pm 0.02$
keV with equivalent width $\sim 100$ eV.

The abundances for O, Ne, Mg, Si, S, Fe were left free to vary, since
these elements have strong lines in the considered wavelength range.
Other $\alpha$-elements with weaker lines (Na, Al, Ar, Ca), which
would not be sufficiently constrained on their own, were bounded
together using solar ratios, accordingly to the proximity in
their atomic number, e.g Na and Al to Mg, Ar and Ca.

\subsection{Multi-temperature plasma}

In the literature, the gaseous emission is usually modelled with one
or two single-temperature components. A `warm' component usually
around 0.7--1 keV is almost ubiquitous \cite{pta99}. A high energy
component is also often reported \cite{dellaceca97,cappi99} in studies
based on ASCA or BeppoSAX data, described either as a power-law, or as
a thermal spectrum with $kT\sim $4--10 keV. Because of the poor spatial
resolution of those satellites it is not clear if this component is
really diffuse or rather due to point sources. Thus, some care should
be taken in comparing literature results with this paper, since in the
following we'll account for both point sources (via the power law
discussed in the previous paragraph) and for hot gas.  A `cold'
component, around 0.1--0.3 keV, is also found sometimes
\cite{dellaceca99}.  In general, the better the quality of data, the
larger the number of components required to provide an acceptable fit.
Thus, the use of multi-temperature models that allow for a fine control
of the DEM, and are also insensitive to the Fe-bias \cite{buote98},
might be preferable but this is hardly feasible, because of demanding
requirements in data quality and computing time.  The grade of our M82
observation is sufficiently high to allow for the first time the use
of multi-temperature models even in spatially-resolved spectroscopy.

\begin{figure*}[t]
  \centering
  \hbox{
  \psfig{figure=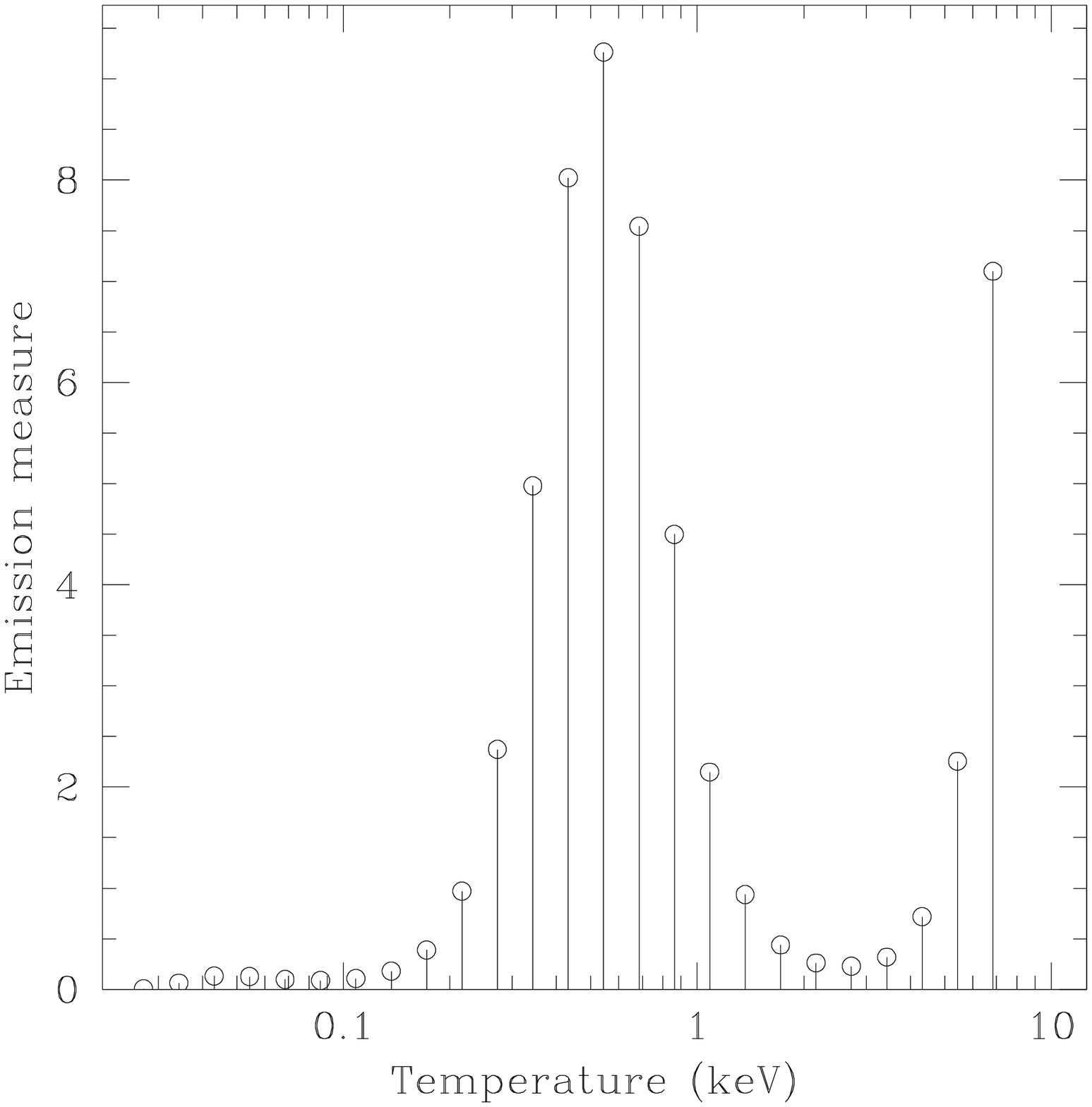,height=.74\columnwidth,width=.82\columnwidth}  
  \psfig{figure=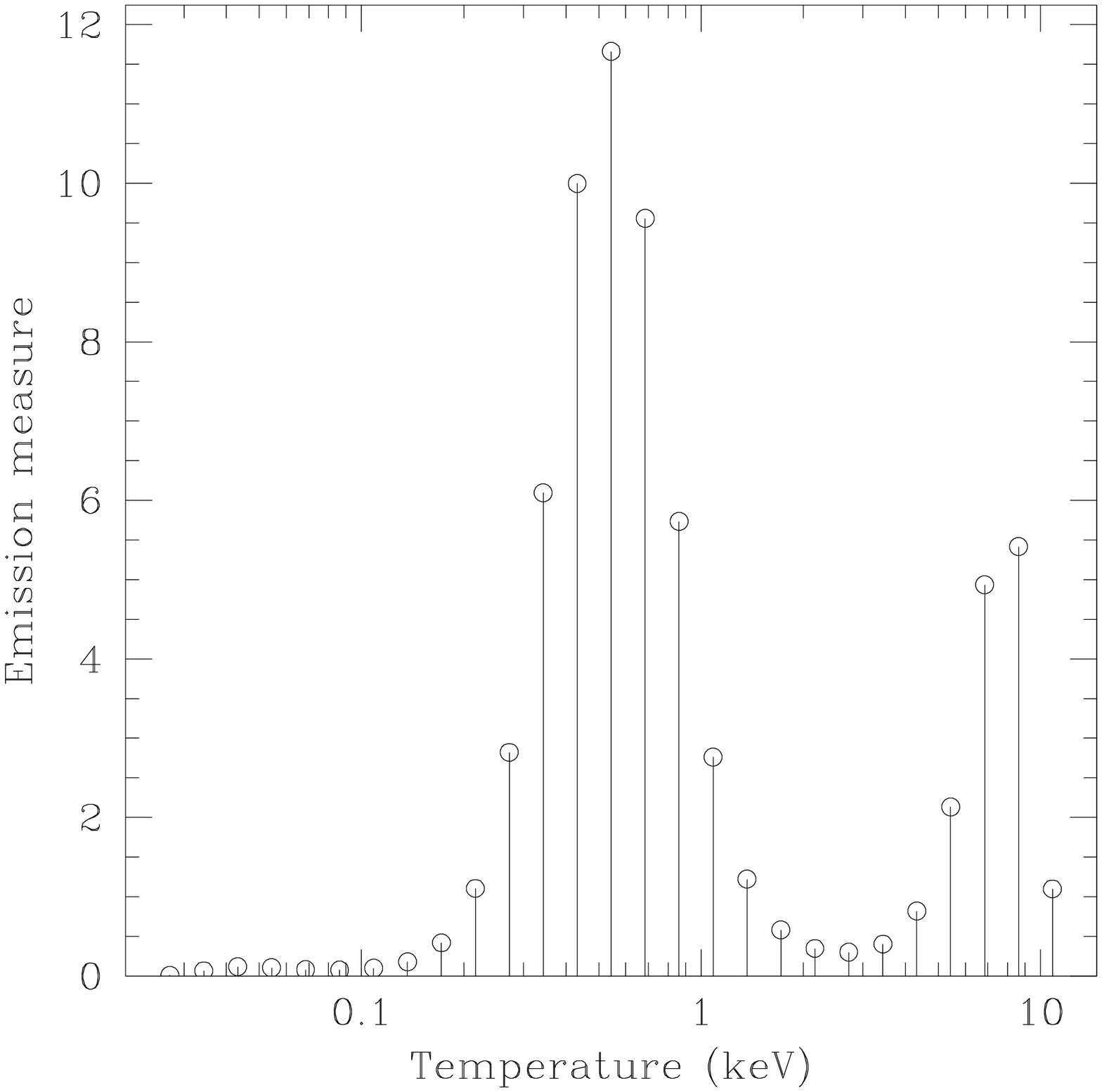,height=.74\columnwidth,width=.82\columnwidth}}
  \caption{The best fit DEM for the central regions of M82 by using
    the standard {\tt c6pvmkl}\ model (left panel) and its
    enlarged-energy-array version (right panel). The total spectrum is
    computed by adding several components, one for each sampling point
    shown here.}
  \label{fig:DEM}
\end{figure*}

The results of the fit showed that a hot component ($kT>\sim 5$ keV)
is required in order to fit the spectrum.  Several versions of the
common MEKAL code \cite{mekal} are present with multi-temperature
flavours in the XSPEC distribution (version 11.3.2g), among which the
most flexible are {\tt c6vmekl}\ and {\tt c6pvmkl}.  These models parameterise
the plasma DEM, i.e.\ the amount of plasma with a given temperature,
with a $6^{\rm th}$ order polynomial, and allow for variable
abundances. The difference between them is that {\tt c6pvmkl}\ restricts
the DEM to positive values, while {\tt c6vmekl}\ also allows negative
values: this means that negative amounts of plasma are permitted. Thus
we chose to use {\tt c6pvmkl}. 

However, these models allow only temperatures with $kT\le 7$ keV,
causing the best-fit DEM to have a sharp cut, clearly noticeable in
Fig.~\ref{fig:DEM} (left panel).  Feeling that this limitation was
somewhat artificial, we modified the available multi-temperature plasma
models by enlarging the allocated energy range, so that the
high-temperature part of the DEM could be better sampled.  The best
fit DEM for both the standard and `enlarged' models are shown in
Fig.~\ref{fig:DEM}.  The improvement in $\chi^2$ was however not
significant, indicating that the DEM sampling was already sufficiently
good: as it may be seen in Fig.~\ref{fig:DEM}, the high-temperature
component is distributed like a rather narrow bell curve, so little
difference is found between a single temperature component and the sum
of two components with very similar temperatures.

%
%
We also checked if the hot component could have a non-thermal origin:
since star forming galaxies are known as radio sources, bremsstrahlung
emission should in principle arise from the same population of
electrons from which the synchrotron emission is originated. Thus we
developed a non-thermal bremsstrahlung model for XSPEC. We found that
the hot thermal component could not be replaced by a non-thermal one,
because of {\em i)} the fit statistics worsens; {\em ii)} given the
observed X-ray flux from the non-thermal bremsstrahlung component, and
hence the required population of free electrons, an unrealistically
low galactic magnetic field would be required to account for the
observed radio (synchrotron) emission.

\subsection{Unidentified lines}

\begin{figure*}
  \centering
  \psfig{figure=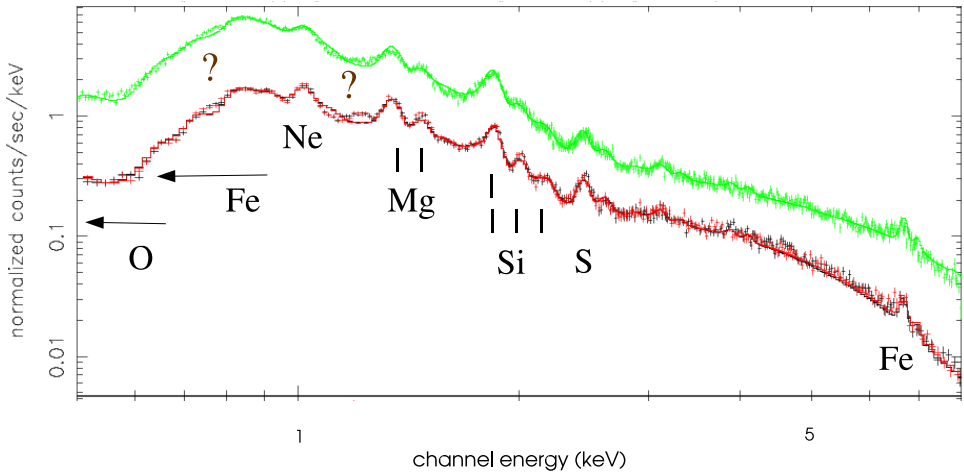,width=.8\textwidth}
  \caption{EPIC spectra of the central ($2^\prime$) region of
    M82. Green points (upper data series): PN. Red and black points
    (lower data series): MOS.  Some of the strongest emission lines
    are indicated. The arrows below Fe and O show line blends. The
    position of the two `unidentified lines' is also shown with a
    question mark.}
  \label{fig:righestrane1}
\end{figure*}

There seems to be two lines in the data which are not present in the
model (see Fig.~\ref{fig:righestrane1}). If the lines are parameterised
with Gaussian models, the best
fit energies are $1.22\pm0.01$ keV 
(equivalent width $\sim 17$ eV) and $\sim 0.78$ keV 
(equivalent width $\sim 20$ eV). 

These lines are present through all the spatial extent of the outflow,
which is about the central $10^\prime$ of the MOS and PN chips.  By
looking at the Chianti line emission database \cite{chianti} the best
candidates for the 1.22 keV line are the Fe XVII line at 1.223 keV
or the Ne X line at 1.211 keV, while for the 0.78 keV line the best
candidate is the Fe XVIII line at 0.7832 keV.  Although {\em i)} the
effective areas in all EPIC cameras in the considered energy range are
a smooth function of the wavelength and do not show any feature around
the considered wavelengths, and {\em ii)} no background feature is present
near the considered wavelengths (moreover, the count rate of the
central region of M82 is about 100 times larger than the background),
we believe that these lines might also be instrumental, background or
model features, for:
\begin{itemize}
\item the 1.22 keV line is flanked at longer wavelengths by an
  absorption-like feature. A close look to Fig.~\ref{fig:righestrane1}
  suggests that a line, if indeed present, should be accompanied by
  problems in modelling the continuum.
\item The 1.22 keV line is not present in the RGS
  data (Fig.~\ref{fig:rgs}), and the other one at 0.78 keV seems to
  be present with lesser intensity.  However, this might also be due to
  the RGS counts being $\sim 1/10$ of the EPIC ones.
\item The line intensity is reduced if the APEC model is used in
  place of Mekal (see below).
\end{itemize}

\begin{figure*}[t]
  \centering
  \psfig{figure=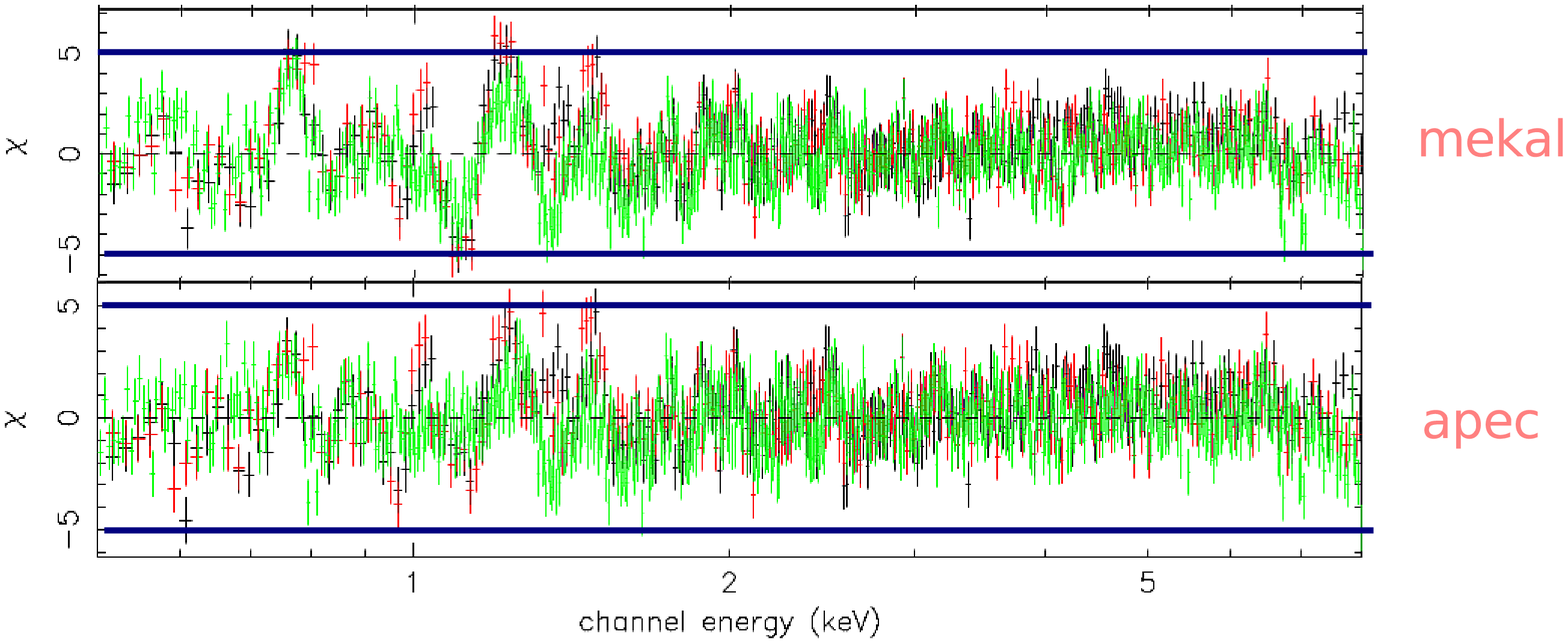,width=.9\textwidth}
  \caption{Residuals of EPIC data with different models. The
  model parameters are best-fit ones and thus are slightly, yet not
  significantly, different in the two cases. The reduced $\chi^2$
  are 1.64 for Mekal and 1.43 for APEC. }
  \label{fig:righestrane2}
\end{figure*}

To further check whether the presence of the two `unidentified
lines' might be related to the details of the spectral model, we
fitted the data using the APEC model \cite{apec}.  Since no
multi-temperature APEC model is present in the current XSPEC
distribution, we built one based on the Mekal one, with the only
difference that the APEC routines are called.  Switching from the
Mekal- to the APEC-based model improves the fit: for the Mekal one,
one gets, considering only the MOS spectra, $\chi^2=1459$ with 889
degrees of freedom (d.o.f.; $\chi^2r=1.64$); for the APEC one,
$\chi^2=1274$  ($\chi^2r=1.43$). A visual inspection of
the residuals confirms that the APEC model provides a better
representation of the data (Fig.~\ref{fig:righestrane2}). As a
technical note, we point out that no significant difference in the
value of $\chi^2$ was obtained for the Mekal model with the {\tt
  switch} parameter set to ``calculate'', as opposed to the default
value of ``interpolate from a pre-calculated table'' (see the XSPEC
documentation).

Thus, we decided to treat these lines as non-astrophysical features. A
detailed comparison of the calculations from the two considered
spectral codes is being performed, and its results are deferred to
the main paper in preparation.

\section{RGS data}

\begin{figure*}[b]
  \centering
  \psfig{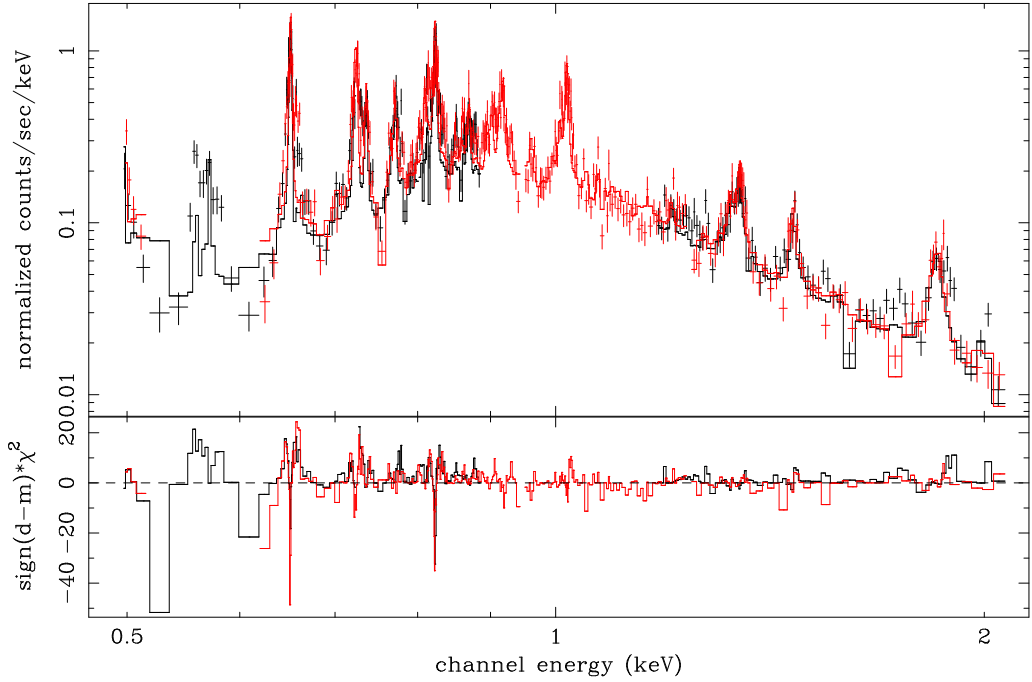}
  \label{fig:rgs}
  \caption{RGS spectrum of M82. Black,
  RGS1; red, RGS2. Only the first order spectra are shown for
  simplicity. The residuals add almost all in the O VII and O VIII lines.  }
\end{figure*}

The RGS spectra for the inner region were extracted with the {\tt
  rgsproc} tool, considering only events within $90\%$ of the PSF
width (i.e.\ $\sim 1^\prime$) in the cross-dispersion
direction. The background spectra were taken from the blank-sky
observations using the {\tt rgsbkgmodel} tool.  The line width in the
RGS data is dominated by the M82 size ($\sim 2^\prime$) so the {\tt
  rgsxsrc} XSPEC model, which convolves the instrumental line width
with the source spatial profile, has been used to account for the
source extent. This model needed to be patched to allow the joint fit
of grating and CCD data. The spectra are shown in Fig.~\ref{fig:rgs}.
The best fit parameters from fits to either the joint EPIC-RGS data
and RGS data alone are consistent with being a spatially-weighted
average of those discussed above (see Fig. \ref{fig:logelem}).

\section{Spatially resolved spectroscopy}
\label{sec:spectro}
We report here on the preliminary analysis conduced on both the northern
and the southern outflows, mainly making use of EPIC.   In order
to study the different properties of the hot gas as it flows and/or is
heated from the central starburst towards the intergalactic space, we
divided both outflows in five sub-regions. Each region has a
rectangular shape, the larger side being parallel to the galaxy major
axis.
The spectra were extracted from
the MOS (0.5--8.0 keV) and PN (1.0--9.0 keV) data, and fitted with the
same model used for the central region, the only difference being
point sources, which in the outflow are sufficiently apart to be
excluded when extracting spectra.

\begin{figure*}[t]
  \centering
  \hbox{ 
    \psfig{figure=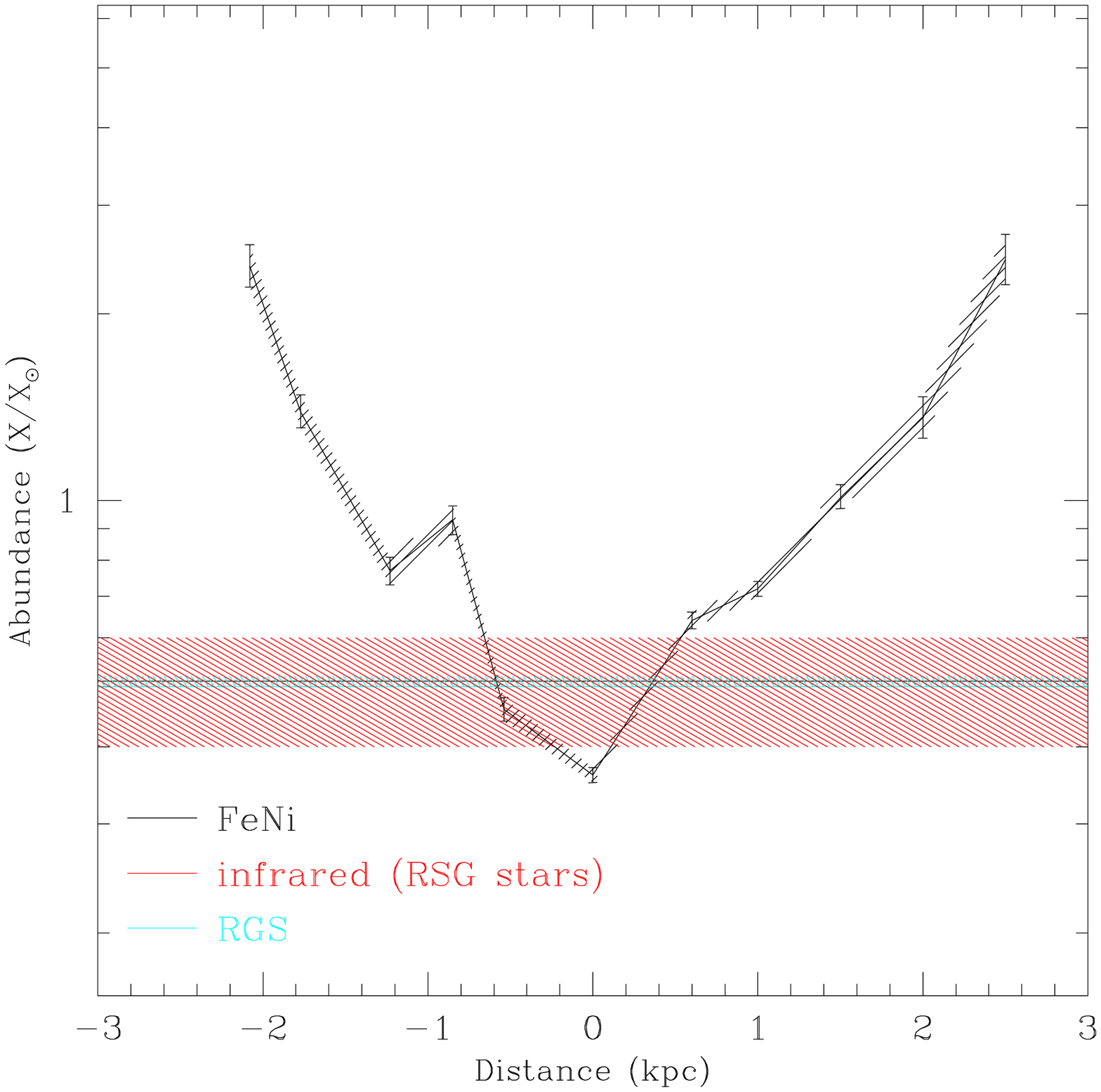,width=.4\textwidth}
    \psfig{figure=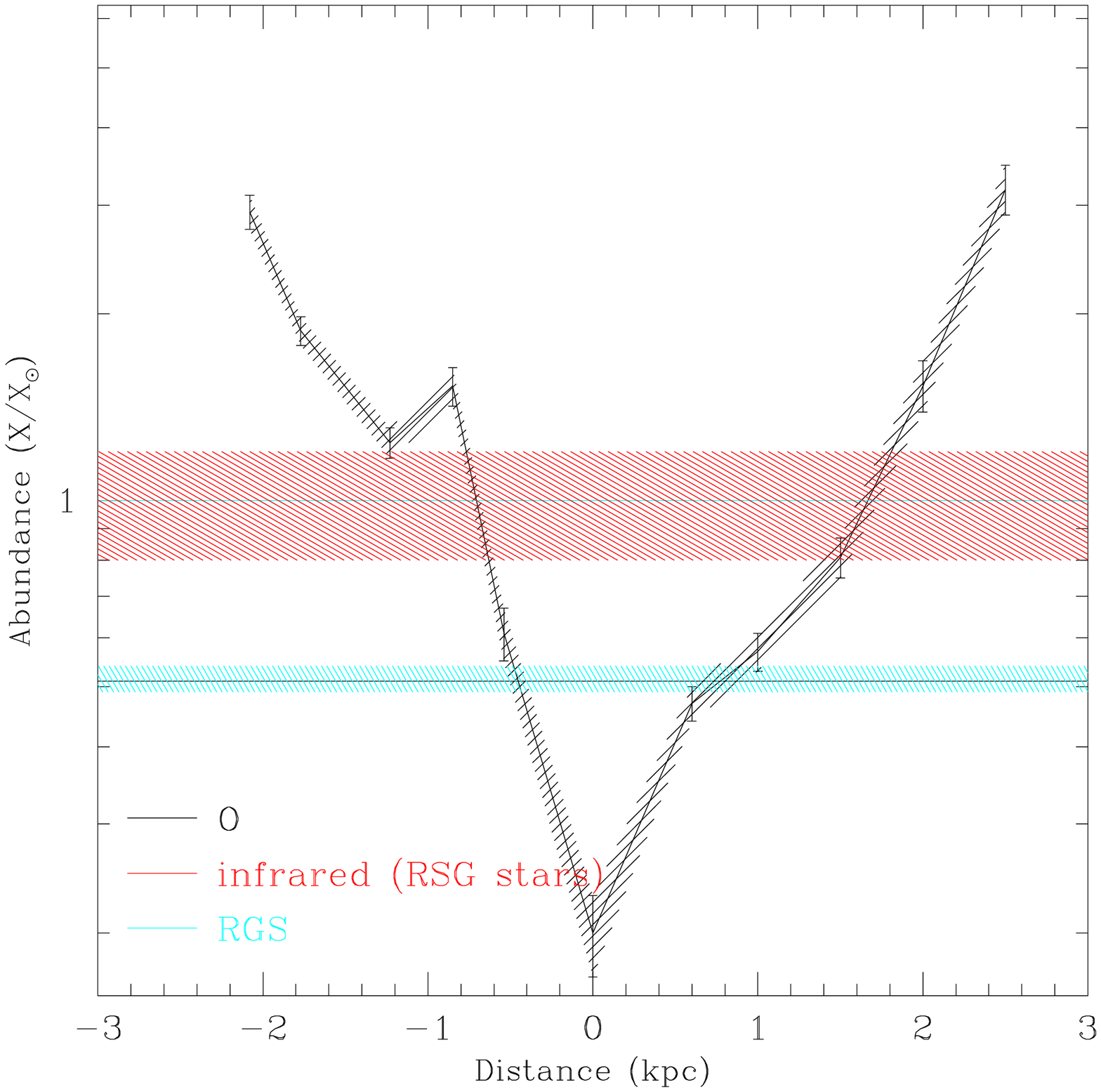,width=.4\textwidth}
  }
  \caption{Variation of chemical abundances with increasing height on
    the galactic plane. Only O and Fe+Ni are shown for brevity. Black:
    abundances from X-ray MOS data.  Red: abundances from infrared data
    (corresponding to red supergiant stars in the galaxy central
    region).  Blue: abundances from joint EPIC-RGS fits.}
  \label{fig:logelem}
\end{figure*}

The best-fit chemical abundances from EPIC and RGS data are shown in
Fig.~\ref{fig:logelem} along with results from our previous paper
\cite{or04} relative to infrared observations for the central
regions.  The abundances are scaled according to the Grevesse \&
Sauval \cite{grev98} solar composition.  No significant changes are
found in the temperature of the plasma. Only one region required in
its spectrum some absorption in excess of the Galactic value.  Our
previous finding, that the inner region of M82 is somewhat devoid of
the lighter $\alpha$-elements, is thus confirmed.  Moreover, it is found that all
elements are more concentrated in the outflow than in the centre. The
centre/outskirt abundance ratio ranges from $\sim 1/10$ for O to $\sim
1/3$ for the heavier elements (Mg, Si, S).

In M82 both the hot gas and the stellar phases trace a very similar Fe
abundance.  Indeed, since Fe is mainly produced by type Ia supernovae
(SN), it is expected to be significantly released in the ISM only
after a few 100 Myr from the local onset of star formation. At
variance, $\alpha$-elements\ (O, Ne, Mg, Si, S) are predominantly
released by type II SN with massive progenitors on much shorter time
scales.

The overall $\alpha$-elements/Fe enhancement is consistent with a
standard chemical evolution scenario only for the heavier elements
(Mg, Si, S).  The lighter elements (O, Ne) show a different
distribution: in the inner regions of the galaxy, the O/Fe and Ne/Fe
ratios appear to be too low for this scenario to hold. In fact, the
X-ray derived O abundance in the centre is less than the
infrared-derived one.

\section{Conclusions}

The analysis of a very deep ($\sim 100$ ks) observation of the
starburst galaxy M82 has opened some new questions, which may be
summarised as:
\begin{itemize}
\item there is a discrepancy in the oxygen abundance between the hot
  gas and stars, which goes in the opposite direction with respect to
  the expectation of chemical evolutionary models;
\item strong vertical abundance gradients in some metals have been
  discovered;
\item the X-ray emitting gas has a ``multi-temperature'' nature.
\end{itemize}

We have worked out some possibilities for interpreting these results,
which will be presented in the forthcoming main paper.

\setcounter{footnote}{-1}  

 

%
%
%

\def\aj{AJ}%
\def\araa{ARA\&A}%
\def\apj{ApJ}%
\def\apjl{ApJ}%
\def\apjs{ApJS}%
\def\ao{Appl.~Opt.}%
\def\apss{Ap\&SS}%
\def\aap{A\&A}%
\def\aapr{A\&A~Rev.}%
\def\aaps{A\&AS}%
\def\azh{AZh}%
\def\baas{BAAS}%
\def\jrasc{JRASC}%
\def\memras{MmRAS}%
\def\mnras{MNRAS}%
\def\pra{Phys.~Rev.~A}%
\def\prb{Phys.~Rev.~B}%
\def\prc{Phys.~Rev.~C}%
\def\prd{Phys.~Rev.~D}%
\def\pre{Phys.~Rev.~E}%
\def\prl{Phys.~Rev.~Lett.}%
\def\pasp{PASP}%
\def\pasj{PASJ}%
\def\qjras{QJRAS}%
\def\skytel{S\&T}%
\def\solphys{Sol.~Phys.}%
\def\sovast{Soviet~Ast.}%
\def\ssr{Space~Sci.~Rev.}%
\def\zap{ZAp}%
\def\nat{Nature}%
\def\iaucirc{IAU~Circ.}%
\def\aplett{Astrophys.~Lett.}%
\def\apspr{Astrophys.~Space~Phys.~Res.}%
\def\bain{Bull.~Astron.~Inst.~Netherlands}%
\def\fcp{Fund.~Cosmic~Phys.}%
\def\gca{Geochim.~Cosmochim.~Acta}%
\def\grl{Geophys.~Res.~Lett.}%
\def\jcp{J.~Chem.~Phys.}%
\def\jgr{J.~Geophys.~Res.}%
\def\jqsrt{J.~Quant.~Spec.~Radiat.~Transf.}%
\def\memsai{Mem.~Soc.~Astron.~Italiana}%
\def\nphysa{Nucl.~Phys.~A}%
\def\physrep{Phys.~Rep.}%
\def\physscr{Phys.~Scr}%
\def\planss{Planet.~Space~Sci.}%
\def\procspie{Proc.~SPIE}%
\let\astap=\aap
\let\apjlett=\apjl
\let\apjsupp=\apjs
\let\applopt=\ao



\end{document}